\magnification=1200
\nopagenumbers
\centerline {\bf {Quantum Fluctuations for Gravitational Impulsive Waves}}
\vskip .7truein
\centerline {Y. Enginer $^{\dagger *}$,
 M.Horta\c csu $^{\dagger *}$, N. \" Ozdemir $^{*}$ }
\vskip .4truein
\centerline { $^{\dagger}$ Physics Dept., T\" UBITAK Marmara Research Center} 
\vskip .1truein
\centerline {Research Institute for Basic Sciences, Gebze, Turkey}
\vskip .3truein
\centerline { $^{*}$ Physics Department, Faculty of Sciences and Letters}
\vskip .1truein
\centerline { ITU 80626 Maslak, Istanbul, Turkey }
\vskip 1.3truein
\centerline { Abstract}
Quantum fluctuations for a massless scalar field in the background
metric of spherical impulsive gravitational waves through Minkowski
and de Sitter spaces are investigated. It is shown that there exist
finite fluctuations for de Sitter space. 

\def\kutu{{\rlap{$\sqcup$}\sqcap}}
PACS Numbers: 98.80 Cq, 11.10-z, 04.62.+v
\baselineskip=18pt
\footline={\centerline {\folio}}
\pageno=1
\vfill\eject
\noindent {\bf{ Introduction}}
\bigskip
The possibility that at the root of the mechanism explaining galaxy 
formation cosmic strings lie is not still ruled out $^{/1}$.  We need 
further experimental data on the anisotropies in the cosmic microwave 
radiation, lensing of quasar images, or gravitational radiation stemming 
from the decay of strings to accept or reject this alternative to 
inflationary quantum fluctuations $^{/1,2,3}$.

However, besides their cosmological implications cosmic strings are also
of interest from the point of quantum field theory in curved
space-time.  Since the space-time generated by a straight cosmic string
possesses a conical structure, it gives rise to finite vacuum fluctuations
of quantized fields.  These effects have been calculated by many authors
$^{4,5}$.  Another effect is seen when the straight cosmic string splits
into two ends.  This results in the emission of a spherical wave.

Exact solutions of Einstein's field equations for spherical impulsive
waves emitted by snapping cosmic strings was found by Nutku and Penrose
$^{/6}$.  It is given by the metric
$$ ds^2= \left( 2dudv+2u^2| d\zeta+{{v\Theta(v)h(\overline{\zeta}
d\overline{\zeta}}\over {2u}} |^2 \right), \eqno (1) $$
where $h$ is the Svhwarzian derivative of an arbitrary function $f(\zeta)$.
$\zeta$ is the stereographic coordinate on the sphere, $u$ is a
Bondi-type luminosity distance and $v$ is a null coordinate which can
be regarded as retarded time.

One may investigate whether the spherical impulsive wave emitted when a 
cosmic string snaps gives rise to finite vacuum fluctuation phenomena.
In Ref. 7 we investigated the vacuum
fluctuations in the background metric of special cases of these solutions.

One can construct similar solutions for impulsive gravitational waves
propagating through the de Sitter universe $^{/8}$. Here one has to simply 
multiply the Minkowski solution by a conformal factor 
$(1+{\Lambda uv \over 6})^{-2} $. Although the Nutku- Penrose solution is 
not conformally related to Minkowski space., an obstruction to the presence 
of finite vacuum fluctuations may sneak in since we are perturbing about the
Minkowski solution.  This new solution in de Sitter space has a parameter,
the cosmological constant $\Lambda$, with the dimensions of mass squared.
In thuis case, perturbation theory is performed about a space where
conformal symmetry is explicitly broken. This allows us to investigate if 
our null results in Ref. 7 persists in the absence of conformal symmetry 
even in the space about which we perform our perturbation expansion.

To search  for vacuum fluctuations, the usual method is the calculation of 
vacuum expectation value,VEV, of the the stress-energy tensor of a scalar 
field in the background metric proposed for the spherical wave $^{/9}$.  
This is achieved by calculating the two point function, $G_{F}(x,x')$ for 
the model as a solution of the equation $(\kutu+\xi R)G_F=-{\delta(x,x')
\over\sqrt{-g}}$ and differentiating the result to obtain 
$<0|T_{\mu \nu}|0>$ after the coincidence limit is taken. The d'Alembertian,
denoted by $\kutu,$ in the above expression is written in the background   
metric of Eq.(1). $R$ is the Ricci scalar and $\xi$ is a constant which
determines the coupling of the field to the metric.

Here we will first perform the calculation by expanding about the Minkowski 
solution.  This means in calculating VEV we will use the vacuum for the 
empty Minkowski space-time with no boundaries. In other words we will define 
our vacuum state in the usual way  by choosing positive frequency modes with 
respect the Killing vector ${{\partial}\over {\partial t}} $ of flat 
Minkowski space-time. Then we will obtain the de Sitter solutions by 
multiplying this solution by conformal factors. We will obtain the de 
Sitter solution by multiplying this solution by conformal factors .

For technical reasons we first take a massive scalar field,
but take the mass equal to zero at the end of the calculation not to 
introduce any mass parameters to the Minkowski calculation. We have more 
control on the consequences of such a parameter when it is introduced via 
the de Sitter universe construction. We use the conformal coupling of the 
scalar field to the metric. This is important in the de Sitter case where 
the Ricci scalar is not zero. In the Minkowski case the Ricci scalar is zero;
so, whether we use minimal or conformal coupling does not matter.

In Ref. 7 we studied a special case of the metric given in Eq. (1)
and used the form  $f=(\zeta)^{1+\delta+i\epsilon}$.  The arbitrary function
in the solution will be the Schwarzian derivative if this function.

We took only the first order terms and found that $h$ is proportional 
to $\delta$ and $\epsilon$.  Since in the Nutku-Penrose solution the 
impulsive wave is generated in a process where a cosmic string snaps, 
we want a small parameter multiplying any possible non zero fluctuation 
effect to act as a signature of the string. Here $\delta$ and $\epsilon$ 
serve this purpose.

In this paper we take a generalized form of this function as
$f=\left( {{A\zeta+B}\over{C\zeta+D}} \right)^{1+i\sqrt{2}\epsilon}$.
We took this form for $f$ , since we knoe that the Schwarzian derivative
of this expression is zero when $\epsilon$ is zero.  
$h$ is proportional to $\epsilon$ in the first order.

Our previous choice for $f$ is a degenerate form of this expression,
where either $A$ and $D$ or $C$ and $B$ are taken to be zero.  We 
take $\delta$ zero in the new calculation since we know from our previous 
work
$^{/7}$ that for fluctuations the case when $\delta$ is not equal
to zero looks essentially the same as the case when $\epsilon$ is not 
to zero. The singular part of the Greens function is the same, only factors
multiplying this term changes;
so, we conclude that taking only one of the parameters finite suffices.

If we agree not to take both $A$ and $C$ in the above expression equal to
zero, we can reduce the new $f$ to the form 
$f=\left( {{\zeta +{B\over A}}\over{\zeta +{D\over C}}} \right)^{1+i \sqrt{2} \epsilon}$ up to an overall
factor in front of this term which can be absorbed in the expansion parameter
$\epsilon$.  
In our specific calculation we take ${B\over {A}} =-1, {D\over {C}} =1$
for convenience.
Since we studied the $A=0$ and the $C=0$ cases seperately, we think that our
new case exhausts all the interesting cases. Whether there are vacuum 
fluctuations or not should not depend upon the location of the zeroes and 
poles of $f$; so, taking the poles at plus and minus one should be able to 
represent the general behaviour of this set of trial functions.  

In the main part of the article, we describe the stress-energy tensor
calculation when
$f=\left( {{\zeta-1}\over{\zeta+1}}\right)^{1+i \sqrt {2} \epsilon}$.
In Section III we apply our results to the de Sitter case. Since the
de Sitter solution is just conformally related to the Minkowski case,
we obtain the de Sitter space Greens functions
from the Minkowski ones just by multiplying
them by conformal factors.  The end results is different, though.
In all the cases studied, we find one non-vanishing component of the
vacuum expectation value of $T_{vv}$
which is proportional to $\delta(v) \Lambda ^2$.  The metric we
used had only one non identically vanishing component for the curvature
tensor which is also proportional to the Dirac delta function of the
variable $v$.The same behaviour is seen in the stress-energy tensor in 
de Sitter universe.
\vfill\eject

\noindent {{\bf{ Calculationg for the Wave Propagating in Minkowski Space}}
\medskip
We start with the Nutku-Penrose metric, given in eq.(1) for
the case where $h$ is
the Schwarzian derivative of
$f=\left({{\zeta-1}\over {\zeta +1}}\right)^{1+i\sqrt{2}\epsilon}$
and treat only first order terms in $\epsilon$.  Here taking complex values
for the exponent corresponds to taking a rotating string, as explained in 
Reference 6.

In the first order in $\epsilon$,$h$ is given by
$$ h={{i8 \epsilon} \over {(x+iy+1)^{2}(x+iy-1)^{2}}} \eqno(2) $$
for $\zeta= {{x+iy}\over {(2)^{1/2}}} $.  This choice gives us the metric
$$ ds^2= 2dudv-(u^2+\epsilon P_1)dx^2 -(u^2-\epsilon P_1)dy^2
+2 \epsilon P_2 dx dy \eqno(3) $$
where
$$ P_1= 64  uv xy {{(x^2-y^2-1)} \over {P^2}} \eqno(4) $$
$$ P_2= 16  uv{{ [ (x^2-y^2-1)^2-4 x^2 y^2]} \over {P^2}} \eqno(5) $$
$$ P = (x^2-y^2-1)^2+4 x^2 y^2.\eqno(6)$$

Just note that for technical purposes we first do the calculation for the 
massive case. Here the mass used can be taken as an infrared parameter.
We take the zero mass limit before the coincidence limit is taken.
We derive the d'Alembertian operator $ {(-g)^{-1/2}} L$, 
written for the Nutku-Penrose metric.  The operator, $L$ , reads, for $v>0$,
$$L= 2u^2 {{\partial^{2}}\over {\partial u \partial v}}
 + 2 u{{\partial } \over {\partial v}}
-{{\partial ^{2} } \over {\partial x^{2}}} - 
{{\partial ^{2}} \over {{\partial y^{2}}}}  + 2m^{2} u^{2}
+{\epsilon \over {u^2}} \left[ P_1( {{\partial ^{2}} \over {\partial x^{2}}} 
- {{ \partial ^{2}} \over {\partial y^{2}}})
-2P_2 {{\partial ^2} \over { \partial x \partial y }} \right]. \eqno(7)$$
We note that the Nutku-Penrose metric does not allow first order derivatives
in $x$ and $y$, for any $h$ used in the original metric in this order.

In order to solve Eq.(7) we use a perturbative scheme. The zeroth order part,
which we call $L_0$, is the flat Minkowski space d'Alembertian function
written in our coordinate system.  Our vacuum will be that of the Minkowski
space with no boundaries.

At this point we may use an expansion of the Feynman Green  Function
using the integral equation approach. Then
$$ G_F= G^0_F+G^0_FVG^0_F+...,$$
where $G^0_F$ is the free Feynman propagator and $V$ is the perturbing part.

We will, instead, start with   the Sturm-Liouville
equation $L\phi _{\lambda} =\lambda \phi_{\lambda} $ and the explicit
construction of the Green function $G_F$ will use the eigenfunctions of this 
equation. $G_F$ will be given by
$$ G_F = - \sum_{\lambda} {{\phi_{\lambda} (x) \phi_{\lambda} ^{*} (x')}
 \over
{\lambda}}.\eqno(8)$$
Here the eigenfunctions of the Sturm-Liouville problem form a complete set.
We perform the calculation to first order in $\epsilon$.  

We expand the operator $L$, the eigenfunction $\phi_{\lambda} $ and the 
eigenvalue ${\lambda}$, in powers of $\epsilon$,
$$ (L_0+\epsilon L_1)(\phi_0+\epsilon \phi_1+\cdots)= 
(\lambda_0+\epsilon\lambda_1+\cdots)(\phi_0+\epsilon\phi_1+\cdots).
\eqno(9)$$
which gives
$$L_0\phi_0=\lambda_0\phi_0,\eqno(10)$$
$$ L_1\phi_0+L_0\phi_1=\lambda_1\phi_0+\lambda_0\phi_1\eqno(11)$$
where
$$ L_0=(2u^2{{\partial^2}\over{\partial u\partial v}}
+2u{{\partial}\over{\partial v}}-{{\partial^2}\over{\partial x^2}}
-{{\partial^2}\over{\partial y^2}}) +2m^{2}u^{2} \eqno(12) $$
$$L_1={1\over{u^2}}\left[P_1({{\partial^2}\over{\partial x^2}}
-{{\partial^2}\over{\partial y^2}})-2P_2{{\partial^2}\over 
{\partial x\partial y}} 
\right].\eqno(13)$$
$L_0$ is the d'Alembertian for the Minkowski case. $P_1$ and $P_2$ are as
defined in Eq.s (4) and (5).

The zeroth order solutions $\phi_0 $ and $\lambda_0$ are found easily .
$$\phi_0={e^{iRv}e^{ik_1 x}e^{ik_2 y}e^{{{-iK}\over{2Ru}}}
e^{{{im^{2} u} \over {R}}}
\over{(2 \pi)^2 u\sqrt{2|R|}}} .\eqno(14)$$
Here $\lambda_0=K-k^2_1-k^2_2$.  $k_1,k_2,R$ and $K$ are constants
definig the different modes.  Note that $\phi_0$  is also a solution in
Minkowski space.

We also find $\lambda_1=(\phi_0,L_1\phi_0)=0$ 
due to the particular form of the operator $L_1$.  

To solve the inhomogenous equation for $\phi_1$, we take
$\phi_1=\phi_0 f.$ 
Then our eq.(11) reduces into $L_1^{'} f=v H $,   
explicitly written as
$$\left[ 2iR u^2 {{\partial }\over { \partial u}}-2i( k_1{{\partial } 
\over {\partial x}}+k_2{{\partial}\over{\partial y}})-{{\partial ^2}\over 
{\partial x^2}}-{{\partial ^2} \over {\partial y^2}}+{2u^{2} 
{{\partial^{2}}\over {\partial u \partial v}}}+{{iK} \over {R}} 
{{\partial } \over {\partial v}}\right] f $$
$$={v \over u} \left[ {{64xy (x^2-y^2-1) (k_1^2-k_2^2)} 
\over {[(x^2-y^2-1)^2+4 x^2 y^2 ]^2}}-32{{k_1 k_2 [ (x^2-y^2-1)^2-4 x^2 y^2]} 
\over {[(x^2-y^2-1)^2+4 x^2 y^2]^2}} \right].\eqno(15)$$

A particular solution of this equation is of the form
$$f= vF_1 (x,y,u)+F_2 (x,y,u).\eqno(16)$$
We find this ansatz by assuming the solution to be of the form
$$f=G_1(v)F_1(x,y,u)+F_2(x,y,u).$$
When this form is substituted in Eq. (15), the form given in Eq.(16) emerges,
with two new coupled differential equations for $F_1$ and $F_2$.
$$ L_2F_1=H, \eqno(17)$$
$$L_2F_2= (-2u^2 {\partial \over {\partial u}} -{iK \over {R}})F_1.
\eqno(18)$$
Here   $H$ is as defined inside the square parenthesis on the
righr hand side of eq. (15) and $L_2$ is given as
$$L_2={2iRu^{2}} {{\partial }\over{\partial u}}-2i(k_1{{\partial} 
\over {\partial x}}+k_2{{\partial}\over{\partial y}}) 
-{{\partial^{2}}\over{\partial x^{2}}}-{{\partial ^{2}}\over{\partial y^{2}}} 
\eqno(19) $$
Since $\phi_1$ is the product of $\phi_0$ with $f$, the boundary conditions
are dictated by $\phi_0$, and the form given by Eq.(16) obeys the boundary
conditions fixed by the zeroth order term, which is the solution in empty
Minkowski space.

At this point we change the variables and define $s={1\over{u}}$.  We use
both variables in our expressions.  We also note that if we use the new
variables $z=x+iy, \overline{z} =x-iy$, 
both the inhomogenous term $H$ and the operator itself seperate.
We find that we get
$$ \left[ -2iR {{\partial} \over {\partial s}} -2i
 \left( (k_1+ik_2) {{ \partial} \over
{ \partial z}} + (k_1-ik_2) {{ \partial } \over {\partial
{\overline {z}} }} \right)
-4{{\partial ^2 } \over {\partial z \partial
{\overline {z}}}} \right] F_1= $$
$$ {-8is } \left[ {{(k_1-ik_2)^2} \over {(\overline {z} -1)^2
 (\overline {z} +1)^2}}
-{{(k_1+ik_2)^2} \over {(z-1)^2 (z+1)^2}} \right].\eqno(20) $$
and a similar equation for $F_2$.

Upon solving the new differential equations, we find particular solutions

$$\eqalign{
F_1=&s\big[2ik_1\big(\tan^{-1}{{2y}\over{x^{2}+y^{2} -1}}
-\big({y \over { (x-1)^2+y^2}}+{y \over { (x+1)^2+y^2}} \big)\big)\cr
-&ik_2 \big(\log {{{(x+1)^2+y^2} \over {(x-1)^2+y^2}}}-
{{2 (x+1)} \over { (x+1)^2+y^2}} 
-{{2(x-1)} \over {(x-1)^2+y^2}}\big)\big]\cr
-&R \big(x\tan^{-1}{2y\over{(x^2+y^2-1)}}
+{{y}\over {2}}\log{(x-1)^2+y^2\over{(x+1)^2 +y^2}}\big)\cr}\eqno(21)$$
and
$$ 
F_2=\left({{iKs} \over {2R}}-1 \right) M_1 (x,y)  +{{iK} 
\over {2(k^2_1+k^2_2)}}(k_1M_2(x,y)-k_2 M_3(x,y)), \eqno(22)$$
where
$$M_1(x,y)=-2y\log\left({{(x+1)^2+y^2}\over{(x-1)^2+y^2}}\right)-2x\tan^{-1}
{{y} \over {x+1}}+2x\tan^{-1}{{y}\over{x-1}}\eqno(23)$$
$$M_2 (x,y) =4xy\log\left({{ (x+1)^2+y^2} \over {(x-1)^2+y^2}}\right) 
+2(x^2-y^2-1)(\tan^{-1}{{y}\over{x+1}}-\tan^{-1}{{y}\over{x-1}})+4y
\eqno(24)$$
$$M_3(x,y)=(2x^2-2y^2-2)\log\left({{(x+1)^2+y^2}\over{(x-1)^2+y^2}}\right)
-4xy(\tan^{-1}{y\over{x+1}}-\tan^{-1}{{y}\over {x-1}})+4x\eqno (25) $$
Now these solutions are used to form the Greens function, $G_F$, of the 
original operator given in eq.(7). Since the zeroth order part of $G_F$ is 
the same as the flat space Greens function, we calculate only the first order 
part.
    
To obtain $G_F$, in first order, we take
$$G_F= {{i\epsilon}\over {2}}
\int_{-\infty}^{\infty}dk_1\int_{-\infty}^{\infty}dk_2
\int_{-\infty}^{\infty}dK\int_{-\infty}^{\infty}{dR\over{|R|}} 
{e^{iR(v-v')}\over{(2\pi)^4 uu'}} 
e^{ik_1 (x-x')}e^{ik_2 (y-y')}e^{{-iK\over{2R}}
\big({1\over{u}}-{1\over{u'}}\big)}$$
$$\times e^{im^{2}{{(u-u')} \over {R}}} 
{\big(\Theta(v)f(x,y,u,v)+ 
\Theta(v')f^{*}(x',y'u',v')\big)\over{K-k_1^2-k_2^2}}. \eqno(26)$$
with $f(x,y,u,v)$ defined in Eq. (16).  Here we use the Schwinger
prescription  to regularize the denominator and use the formula
${{1}\over {A-i\delta}}= i\int_{0}^{\infty} e^{-i\alpha A -\delta
\alpha} d\alpha$ where $\delta$ is a positive constant approaching zero.
The infinite integration over $k_1,k_2$ is regularized by the denominator.
The result is the Feynman propagator with its usual contour.

We need $G_F$ to calculate the vacuum expectation value of the stress-energy 
tensor $T_{\mu \nu}$. We are particularly interested in whether $<T_{vv}>$
has a finite part. Since in $T_{vv}$ $x$ or $y$ derivatives do not exist, 
we take the coincidence limit in $x$ and $y$ at this stage.  This puts terms  
of the type $ (x-x')^2$ and $(y-y')^2 $ to zero. Such terms either multiply 
the expressions obtained, or appear in the `geodesic distance' term
$\sigma= (u-u')(v-v') -{{uu'}\over {2}} ((x-x')^2+(y-y')^2),$
which appears in the denominator. Taking the coincidence limit in $x$ and 
$y$ only simplifies the calculation at this stage and does not affect the 
end result.

When we take $x$ equals $x'$ and $y$ equals $y'$, all the terms that are odd
in $k_1$ and $k_2$ vanish. The remaining terms are given by
$$G_F(s,s',v,v',x,y)= 
{{-i\epsilon}\over{uu^\prime 2(2\pi)^4}}
\int{dR\over{2\vert R\vert}}\int dK\int dk_1\int dk_2\int_{0}^{\infty}d
\alpha e^{{-iK\over{2R}}\big({1\over{u}}-{1\over{u'}}\big)}$$
$$\times e^{iR(v-v')} e^{im^{2}{{(u-u')}\over{R}}}
e^{-i\alpha\big(K-k^{2}_1 -k^{2}_2 \big)-\alpha \delta}$$
$$\times\big[\left({{K}\over{2R}}
(s\Theta(v)-s'\Theta(v'))+i(\Theta(v)+\Theta(v')) 
\right)M_1 (x,y)
+{2R M_4(x,y)}(v\Theta(v)-v'\Theta(v'))\big]\eqno(28)$$ 
where 
$$M_4(x,y)={1\over 2}y\log{{(x+1)^2+y^2}\over{(x-1)^2+y^2}}
-x\tan ^{-1}{{2y}\over{ x^2+y^2-1}},\eqno(29)$$ 
and $M_1 (x,y)$ was defined in Eq. (23).

We perform the integrations over $k_1,k_2,K$ and $R$. The result of the 
$\alpha$ integration can be shown to result in Hankel functions which 
degenerates into a monomial when the $m$ going to zero limit is taken.  
Hence for the massless field case the Greens Function $G_F$ reads
$$G_F=\left({\epsilon\over{16\pi}}\right)\big[{1\over{(u-u')(v-v')}}
\big( \left({{(u\Theta(v')-u'\Theta(v))}\over{(u-u')}}-
(\Theta(v)+\Theta(v'))\right) M_1(x,y)$$ 
$$+M_4(x,y){{(v\Theta(v)-v'\Theta(v'))}\over{(v-v')}} \big) \big]\eqno(30)$$

We note that the singularity structure in the coincidence limit of this
expression  displays exactly that of the free Minkowski Green Function.
There is just a modulating factor which is finite in the coincidence limit.
Since our perturbative solution $\phi_1$ is given as a factor times the
Minkowski solution, the singularity structure of the Minkowski vacuum is
carried over this solution.  This part may be regularized by renormalizing
the free parameters of the theory.

The empty space Green function  goes as ${{1}\over {4{\pi}^2 \sigma}}$.
Here $\sigma$ is the square of the geodesic distance between the two points
given by Eq. (27).  A well known theorem $^{/10}$ states that if the
metric is of the Minkowskian shape for a certain region of space-time,
the Green Function will be of the Hadamard type $^{/11}$,
$$G_F={{A}\over {\sigma}} +B log \sigma +C ,$$
where A,B,C are finite quantities at the coincidence limit; $\sigma$
is the geodesic distance between the points of $G_F$.  Our hope, in this
calculation, was to find a finite term $C$ given in the above formula.
Unfortunately our end had $B$ and $C$ identically zero.  We choose our
boundary conditions of the renormalized Green function of the renormalized
Greens function so that we get zero as $u$ and $v$ go to infinity.
This condition fixes $C_0$ in the expansion $C= \Sigma C_l \sigma_l$,
therefore completely fix $C$ uniquely equal to zero.

Another way of obtaining this result will be applying the Adler-Lieberman-Ng
prescription $^{/12}$ to this problem.  An ambiguity may arise only by a
local conserved tensor which comes out to be proportional to $a_2$ of the
adiabatic expansion in the DeWitt-Schwinger method $^{/13}$.  This term is
identically zero in our case; so, even this possible ambiguity is not
present.  Other references to the uniqueness property of the VEV of the
stress-energy tensor can be found in Ref. 14.
\bigskip
\noindent
{\bf{ Calculation for the Wave Propagating in de Sitter Universe} }
\bigskip
We conjecture that we have to introduce a mass parameter to our space
in order to get non zero result for the vacuum fluctuations if we insist 
performing our calculation perturbatively.  One can check whether this is 
true by doing the similar calculation in de Sitter universe.  Indeed 
we find that our end result changes if we look at the spherical
impulsive gravitational waves propagating through the de Sitter universe.  
It is shown that $^{/ 8}$ an impulsive spherical gravitational wave solution 
exists in de Sitter space. Here one has to simply multiply the Minkowski 
solution  by a conformal factor 
$(1+{{\Lambda uv}\over{6}})$ where $\Lambda$ is the cosmological constant 
which comes out proportional to Ricci scalar in this metric.
The metric  now reads
$$ds^{2}=(1+{{\Lambda uv}\over{6}})^{2}\left(2dudv+2u^{2}|d\zeta+{{v\Theta(v) 
h(\overline{\zeta})d\overline\zeta}\over{2u}}|^{2}\right)\eqno(31)$$
We take the same $h$ as before, as given in eq.(2).   Since we use conformal
coupling, the operator $L$ is modified to include also the Ricci scalar term 
for this metric which is no longer zero.

We know from general arguments that $^{/9}$
$$G_F^{S}=(1+{{\Lambda uv}\over{6}})G_F^{M}(x,x')(1+{{\Lambda u'v'}\over{6}}) 
\eqno(32)$$
where $G^{S}_{F}$ and $G_{F}^{M}$  are the de Sitter and Minkowski space
Greens functions respectively.

Note that
$$(1+{{\Lambda uv}\over{6}})(1+{{\Lambda u'v'}\over{6}})=
(1+{{\Lambda UV}\over {6}})^{2} + {{\Lambda} \over {12}} (u-u')(v-v') $$
$$+({{\Lambda} \over {12}})^{2}\left(-(u-u')^{2}V^{2} -(v-v')^{2}U^{2}
+{{(u-u')^{2}(v-v')^{2}}\over{4}}\right)\eqno(33)$$
where $U={{u+u'} \over {2}}, V={{v+v'} \over {2}} $.
Using this expansion we see that for $G_F^{S}$ we get  terms in the 
coincidence limit that give finite terms to $<T_{vv}>$
$$\lim _{u\to u^\prime,v\to v^\prime}
{\partial^2\over{\partial v \partial v'}}
\left[{{\Lambda^{2}(u-u')^{2}(v-v')^2}\over{(24)^{2}}}G^{M}_{F}(x,x')
\right]$$
$$={-u\epsilon\Lambda^{2}\over{(96)^2\pi}}
\left(2x\tan^{-1}{{y}\over{x+1}}-2x \tan^{-1}{y\over{x-1}}
-2y\log\left({(x-1)^2+y^2\over{(x+1)^2+y^2}}\right)\right)
\delta(v)\eqno(34)$$

We find that in first order calculation for spherical impulsive waves 
propagating through the de Sitter universe, we get one non-zero component
of the stress-energy tensor.

If we perform the similar calculation for the cases studied in Ref.7,  
we get nonzero contribution, given in the same way. Here $h$ in eq.(1)  
comes from $ f= (\zeta)^{1+ \delta+i\epsilon}$. 
Using the results given in Ref.7, we get
$$T_{vv}={-\Lambda^2 u\delta(v)\over{2(96)^{2} \pi}} 
\left(\delta\log(x^2+y^2)-4\epsilon\tan^{-1}{y\over x}\right).\eqno(35)$$
This result is finite, contrary to the case for propagating in Minkowski
space.
\bigskip
\noindent
{\bf{Conclusion}}
\bigskip
Here  we calculated the stress-energy tensor of a scalar field in the
space-time  of a spherical impulsive gravitational wave propagating 
through the Minkowski and de Sitter universes.  

We verified the common knowledge that fluctuations which are null in 
Minkowski space may become finite in de Sitter space.  We found that VEV of 
one component of the energy momentum tensor in first order is proportional  
to $\delta(v)$ which is the signature of the impulsive wave solution, in de 
Sitter space.  This is the same factor to which the only nonzero component of 
the curvature tensor is also proportional.

Our results are found in first order perturbation theory, but are proportional 
to the square of the curvature scalar. They are an example of extracting non 
trivial information out of first order perturbation theory.

As a technical remark note that first order solutions to the Sturm Liouville 
problem arising from this metric can be expressed in terms of sum of 
holomorphic and antiholomorphic functions.  This occurs after we factorize
the solution into two parts, one part proportional to the flat space solution.
This fact simplifies our calculation, since then second order differential 
equation decomposes in to a pair of first order equations which are easily 
integrated. In our previous papers $^{/7}$, we used brute force Greens 
function method to integrate the inhomogenous equation. Since we have to 
perform integrals where the integration range is infinite, there  were 
points where regularizations were used. In the new method we do not need 
any  regularizations in calculating   $G_F$. The result is free from any 
possible ambigiuties the regularization may cause.

We reduce first order eigenfunction calculation to the solution of a
couple of first order differential equations. This is possible since both
the inhomogenous term and the differential operator decomposes into a sum
of two terms if a set of variables are used.
\bigskip
{\bf{ Ackowledgement:}} We thank Prof. Dr. Yavuz Nutku for leading us
to this general class of problems
for many interesting discussions.  This work is partially supported by 
through TBAG \c CG.1 by T\" UBITAK , the Scientific and Technical Research 
Council of Turkey. Y.E. also acknowledges a scholarship given by T\" UBITAK.
M.H. is also supported by TUBA, the Academy of Sciences of Turkey.                   
\vfill\eject
\noindent
{{\bf{REFERENCES}}

\item {1.}   A.Vilenkin and E.P.S. Shellard,
{\it{ Cosmic Strings and other Topological Defects}}, 
Cambridge University Press, Cambridge (1994);

M.B. Hindmarsh and T.W.B. Kibble, Reports on Progress in Physics, 58
(1995) 477.

\item {2.}  L. Perivolaropoulos, " Cosmic String Theory: The Current Status",
MIT preprint, MIT-CTP-2375 (1994).

\item {3.}   Ed. by G.W. Gibbons, S.W. Hawking and T. Vachaspati,
{\it{ The Formation and
Evolution of Cosmic Strings}}, Cambridge University Press, Cambridge (1990).

\item {4.} T. Vachaspati and A. Vilenkin , Phys. Rev. D31 (1985) 3052.

\item {5.} L. Parker , Phys. Rev. Lett. 59 (1987) 1365;
 A. C. Smith, in reference 3, p.263;
V. Sahni, Modern Phys. Lett. A 3 (1988) 1425;
 T.M. Helliwell and K.A. Konkowski, Phys. Rev. D34 (1986) 1908;
B. Linet, Phys. Rev. D33 (1986), D35 (1987) 536.

\item {6.} Y. Nutku and R. Penrose, Twistor Newsletter 34 (1992) 9.

\item {7.} M.Horta\c csu, J. Math. Phys., 34 (1993) 690;
 M. Horta\c csu and R. Kaya, J. Math. Phys., 35 (1994) 3043.

\item {8.} P.A. Hogan, Phys. Lett. A171 (1992) 21.

\item {9.} N.D. Birrell and P.C.W. Davies, {\it{ Quantum Fields in Curved 
Space}}, Cambridge University Press, Cambridge (1982).

\item {10.} S.A.Fulling, M. Sweeny, R.M.Wald, Comm. Math. Phys. 63 (1978) 
257.

\item {11.} J. Hadamard, {\it{ Lectures on Cauchy's problem in linear partial
differential equations}} (Yale University Press,1923).

\item {12.} S. L. Adler, J. Lieberman, Y. J. Ng, Ann. Phys (N.Y.) 106 (1977) 
279; S. L. Adler, J. Lieberman, Ann. Phys. 113 (1978) 294; R. M. Wald, 
Phys. Rev. D17 (1978) 1477.

\item {13.} J.Schwinger, Phys. Rev. 82 (1951) 664; B.S. DeWitt, {\it{
The dynamical Theory of Groups and Fields}},(Gordon and Breach, New York,
1965); B.S.DeWitt, Phys. Reports 19C (1975) 297.

\item{14.} R.M.Wald, Comm. Math. Phys. 54 (1977)1; Stephen A. Fulling,
{\it{ Aspects of Quantum Field Theory in Curved Space-Time}}, ( Cambridge
University Press, Cambridge 1989), especially Chapter 9 on Renormalization;
a summary is found in Chapters 6 and 7 of reference 9.

\end